\documentclass[12pt,preprint]{aastex}
\usepackage{graphicx}

\newcommand{\yim}{Y_{\rm ini}}
\newcommand{\ysm}{Y_{\rm surf}}
\newcommand{\yims}{Y_\odot^{\rm ini}}
\newcommand{\ysms}{Y_\odot^{{\rm surf}}}

\newcommand{\ysssmp}{$Y_{\rm surf,0}$}
\newcommand{\yissm}{$Y_{\rm ini,0}\ $}
\newcommand{\ysssm}{$Y_{\rm surf,0}\ $}

\shorttitle{Solar initial helium abundance}
\shortauthors{Serenelli \& Basu}

\begin{document}

\title{Determining the initial helium abundance of the Sun}

\author{Aldo M. Serenelli\altaffilmark{1}}
\affil{Max  Planck Institute  for  Astrophysics, Karl  Schwarzschild Str.   1,
  Garching, D-85471, Germany}
\email{aldos@mpa-garching.mpg.de}
\and 
\author{Sarbani Basu}
\affil{Department of  Astronomy, Yale University, P.O. Box  208101, New Haven,
  CT 06520-8101, USA}
\email{sarbani.basu@yale.edu}

\altaffiltext{1}{Instituto de 
    Ciencias  del Espacio  (CSIC-IEEC),  Facultad de  Ci\`encies, Campus  UAB,
    08193, Bellaterra, Spain}

\begin{abstract}
We determine the dependence of the initial helium abundance and the present-day
helium abundance in the convective envelope of solar models ($\yim$ and $\ysm$
respectively) on the  parameters that are used to construct  the models. We do
so  by  using  reference  standard  solar  models  to  compute  the  power-law
coefficients of the dependence of $\yim$ and $\ysm$ on the input parameters.
We use these dependencies to determine the correlation between $\yim$ and
$\ysm$ and use this correlation
to eliminate uncertainties in $\yim$ from all 
solar model input parameters except the microscopic diffusion rate. We find an
expression for $\yim$ that depends only on $\ysm$ and the diffusion rate. By
adopting the helioseismic determination of solar surface helium abundance,
$\ysms= 0.2485\pm0.0035$, and an uncertainty
of 20\% for the diffusion rate, we find that the initial solar
helium abundance, $\yims$, is $0.278 \pm 0.006$ independently
of the reference standard solar models (and particularly on adopted the solar abundances)
used  in the derivation  of the  correlation between  $\yim$ and  $\ysm$. When
non-standard solar models with extra mixing  are used, then we derive $\yims =
0.273 \pm 0.006$. In both cases, the derived $\yims$ value is higher than that
directly derived  from solar model  calibrations when the low  metalicity solar
abundances (e.g. by Asplund et al.) are adopted in the models. 
\end{abstract}

\keywords{Sun:  abundances ---  diffusion  --- Sun:  helioseismology ---  Sun:
  interior} 

\section{INTRODUCTION}

The primordial helium abundance, $Y_{\rm P}$, can be determined to a high
accuracy thanks to the measurement of
cosmological   parameters,   particularly   of  the   baryon-to-photon   ratio
$\eta_{10}$, by WMAP 
\citep{wmap:2007}. The most recent WMAP results 
\citep{wmap:2010} give $Y_{\rm P}=0.2486 \pm 0.0006$ (using the analytic 
formula presented by \citealt{steigman:2007} relating $Y_{\rm P}$ and $\eta_{10}$),
but rely on the assumption 
of Standard Big Bang Nucleosynthesis (SBBN). On the other hand, empirical
determinations of $Y_{\rm P}$, that can be used to test the SBBN scenario,
rely on the determination of helium abundances in metal poor HII regions and
extrapolations to zero metalicity, assuming a linear relation between
helium and metalicity (or oxygen) with a given slope $\Delta Y / \Delta Z$ (see
\citealt{peimbert:2007} for a detailed account of the method and uncertainties
involved). The results are not exempt of controversy. For example,
\citet{peimbert:2007} find $Y_{\rm P}=0.2477 \pm 0.0029$, in excellent agreement
with WMAP+SBBN results, while other authors (e.g., \citealt{izotov:2010}), find
higher values, $Y_{\rm P}=0.2565 \pm
0.0010 \, \mathrm{(stat)}\pm 0.0050 \, \mathrm{(syst)}$, and suggest deviations
from SBBN.  Discrepancies are not confined  to   $Y_{\rm P}$,
but  are also  found in  the given  values  of $\Delta  Y /  \Delta Z$.  Other
attempts to determine 
$\Delta Y / \Delta Z$, 
both observationally and theoretically, include the use of the broadening of
the lower main sequence (see e.g. \citealt{casagrande:2007}
and \citealt{gennaro:2010} for recent works), the study of galactic stellar
populations \citep{renzini:1994}, modeling eclipsing binaries
\citep{ribas:2000}, construction of galactic chemical evolution models
(\citealp{timmes:1995,chiappini:2003,carigi:2008}). The 
values obtained for $\Delta Y / \Delta Z$ range approximately between 1.4 and
2.5 for different authors and methods. 

The initial composition of stars determines their structure and evolution and,
by extension, the characteristics of stellar populations. The initial helium
abundance of stars is in most cases impossible to determine directly, hence
indirect methods have to be used. A common approach used in stellar and
stellar populations modeling is to assume a
value for $\Delta Y / \Delta Z$ and use the solar helium and metalicity as an
anchor point for the relation. This is a routine procedure used in, for example,
libraries of stellar evolutionary tracks to define the initial composition used
in the stellar models (e.g.,
\citealp{schaller:1992,yy2:2001,pietrinferni:2004})
The initial solar helium abundance, however, can
not be determined directly. It is well established that element diffusion
takes place in the Sun and that the present-day photospheric abundances of helium
and metals are lower than the initial ones. Therefore, the initial solar helium
abundance is usually obtained from calibrating solar models, which then introduces
model dependencies on the determined values. An illustrative example is
the effect of solar photospheric metalicity determinations on the
inferred initial solar helium abundance. Solar models that use older 
solar abundances like those of \citet{gn93} or \citet{gs98} lead to models with
initial helium values in the models in the range $\yim \approx 0.272 - 0.275$
\citep{bpb:01,booth:03}, while models that incorporate newer determinations of
solar metalicity \citep[e.g.]{ags05,lodders:2009,agss09} give $\yim \approx
0.260 - 0.263$ \citep{bs05,guzik:2005,ssm09}. While the 
models with the older abundances have present-day surface helium abundances
$\ysm$ consistent with helioseismic determinations of the 
helioseismology determinations of the present-day solar convection-zone
helium abundance, the models with the lower metal abundances do not.
It is interesting to note also that if we adopt $\yim \approx
0.260 - 0.263$ values together with initial solar metalicities
derived from the same solar models, i.e. $Z_{\rm ini} \approx 0.014 - 0.015$, then
values as low as 0.8 are obtained for $\Delta Y / \Delta Z$ (assuming $Y_{\rm
P}$ from  SBBN). This is in  disagreement with all  accepted determinations of
$\Delta Y / \Delta Z$ as we have summarized above.

In this work, we propose a simple method to determine the initial solar helium
abundance based on the use of solar models to calibrate the relation between
the initial and the present-day surface helium abundances. Although derived
with the help  of solar models, we then show that  the predicted initial solar
helium abundance ($\yims$; not to be confused by the initial helium abundance
$\yim$ resulting from solar model calibrations) is a robust result, i.e. a
very wide range of solar models (standard and non-standard) lead to the same
$\yims$ value. The structure of the paper is as follows: in
\S~\ref{sec:powerlaws} we determine
the effect of changes in the input parameters used in  solar models (including
metal abundances, diffusion rate, nuclear cross sections, etc.) on the initial
and present-day helium abundances of the solar models; in
\S~\ref{sec:correl} we derive a simple analytic relation between $\yim$ and
$\ysm$ that minimizes the uncertainties originating in the inputs to solar
model calculations; in \S~\ref{sec:results} we present our main results and
discuss their range of applicability; finally, we summarize our findings and
conclusions in \S~\ref{sec:summary}.

\section{POWER-LAW DEPENDENCE OF $\ysm$ AND $\yim$}
\label{sec:powerlaws}

For this paper, we use the same technique that has been used earlier
for analyzing neutrino fluxes \citep[see Chap.7]{jnbbook:1989}.
We express the dependence of
the  initial ($\yim$)  and present-day  ($\ysm$) helium  abundances of solar
models as a product of powers of the different input parameters:

\begin{equation}
\left( \frac{\yim}{Y_{\rm ini,0}} \right)= \prod_{i=1,N}
\left(\frac{p_i}{p_{i,0}} \right)^{\alpha_i} 
\label{eq:yini}
\end{equation}

\begin{equation}
\left( \frac{\ysm}{Y_{\rm surf,0}} \right)= \prod_{i=1,N} 
\left(\frac{p_i}{p_{i,0}} \right)^{\beta_i}.
\label{eq:ys}
\end{equation}

In  these equations,  the input  parameters in  solar  models (nuclear
cross sections, abundances of individual elements, age, etc.) are represented
by  a  set  of  parameters  $\left\{p_i\right\}_{i=1,  N}$  with  central,  or
reference, values denoted with the sub-index ``0''. These parameters comprise: 
metal  abundances,  nuclear  reaction  rates, element  diffusion  rate,  solar
surface parameters (luminosity, age); a detailed list is given in
Table~\ref{tab:pwl}.   The  exponents  $\alpha_i$ and  $\beta_i$  respectively
determine the  dependence of $\yim$ and  $\ysm$ on the  input parameters. They
are defined as 
\begin{equation}
\alpha_i\equiv \frac{\partial \log{\yim}}{\partial \log{p_i}}, \ \ \ \ 
\beta_i\equiv \frac{\partial \log{\ysm}}{\partial \log{p_i}}.
\label{eq:index}
\end{equation}

\begin{figure}
\includegraphics[scale=0.7]{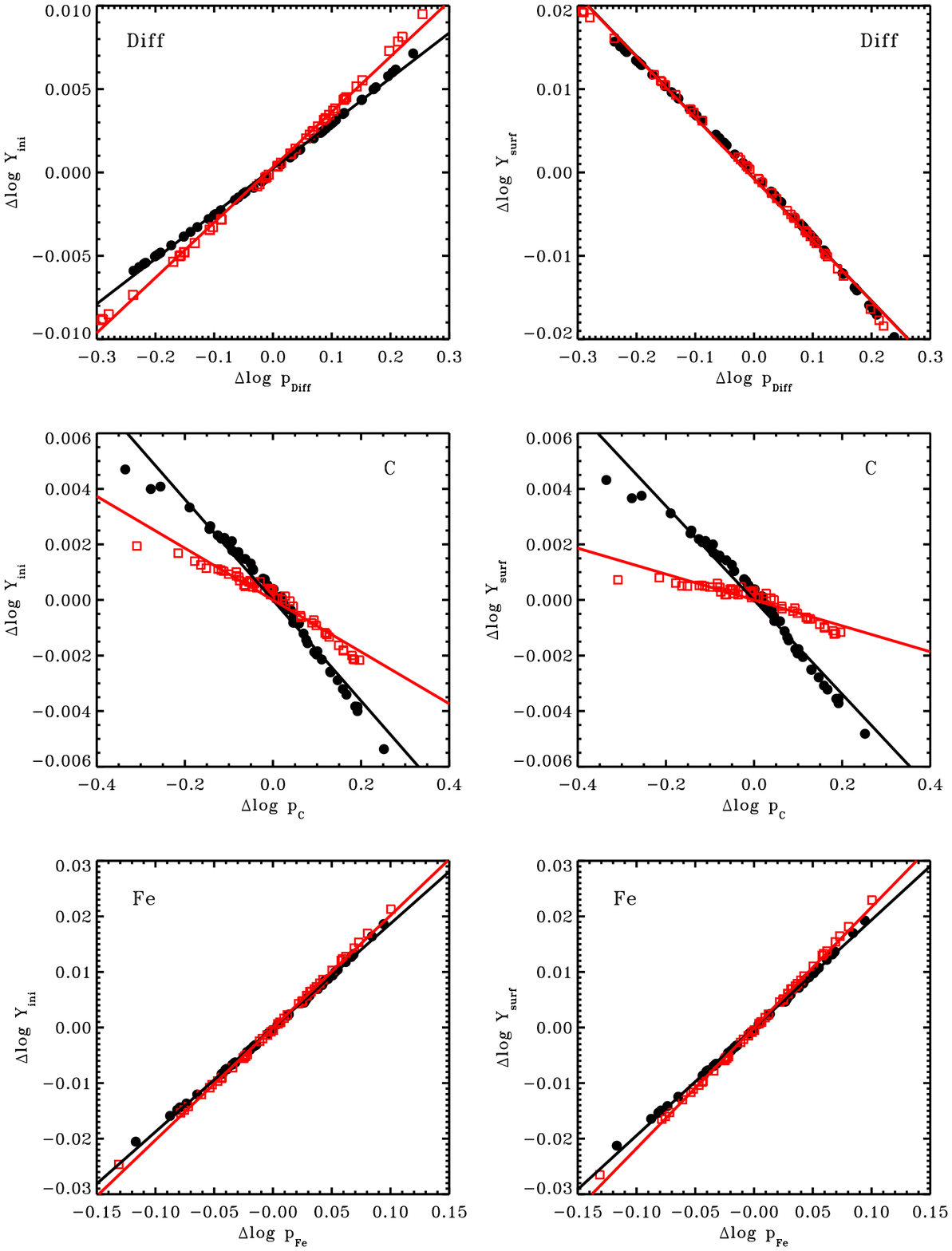}
\caption{Some examples of the derivation of power-law exponents. Left panels
show results for $\yim$ and right panels for $\ysm$. From top to bottom,
results correspond to changes in the diffusion rate, carbon, and iron
abundances. Black filled circles: results for GS98 composition; red empty
squares: results for AGS05 composition. In each
panel, lines denote the linear fits to the model results; the slope is
the power-law exponent (see Equations~\ref{eq:index}). 
\label{fig:fits}}
\end{figure}

The indices  in Equations~(\ref{eq:index}) were calculated for
standard  solar  models  (SSM)   using different determinations of
photospheric solar composition. We chose the compilations by
\citet{gs98} and \citet{ags05}, hereafter GS98 and AGS05 respectively, and
{\em conservative uncertainties}, according to the definition by
\citet[hereafter BSB06]{mcpaper}. A discussion of the central values for
all parameters, as well as their uncertainties, is presented in that work.
To derive the indices we proceeded as follows: for each input parameter
$p_i$ we computed a set of 50 SSMs where the value of $p_i$ is drawn
from a Gaussian distribution, with standard deviation equal to the assumed
uncertainty of the parameter, while all other input parameters are kept fixed
and equal to their central value. The logarithmic derivatives defined
in Equations~\ref{eq:index} are then obtained from fits to each of the small
Monte Carlo simulations described above. Some examples are shown in
Figure~\ref{fig:fits}.
The values of these dependencies are tabulated  in Table~\ref{tab:pwl}. 
The exponents derived  for both GS98 and AGS05 solar  models are similar. This
is so because relative changes in input parameters produce relative changes in
solar  model  properties that  do  not,  to the  first  order,  depend on  the
reference solar model. Even more important for this work is that the ratios 
$\alpha_i/\beta_i$ are almost independent of the reference solar models used. 
Differences in the values obtained for
GS98 and AGS05  models  will  be used  later  as  a  way  to determine  the 
systematic uncertainty of the method presented in this work.

Radiative opacities can not be represented by a single parameter. Hence to
account for the influence of opacities on $\yim$ and $\ysm$ we have proceeded
in the following way: for both AGS05 and GS98 compositions we have computed
two solar models, one with OPAL opacities and the other one with OP opacities,
while keeping all other input physics the same in both models. For each
composition we define the
$1\sigma$ uncertainties in $\yim$ and $\ysm$ as half the
difference in the  values obtained for models with the same composition
but the two different opacity sources. 
We find that for $\ysm$ the uncertainty from opacity is
0.23\%(0.20\%), while for $\yim$ is 0.30\%(0.28\%) for the AGS05(GS98)
composition. The uncertainties between $\yim$ and $\ysm$ are found to be fully
correlated. We account for opacity uncertainties by adding Gaussian dispersions
with the above defined standard deviations to the distributions of $\yim$ and
$\ysm$ that we  generate using Equations~\ref{eq:yini}~and~\ref{eq:ys}.

\begin{figure}
\includegraphics[scale=.8]{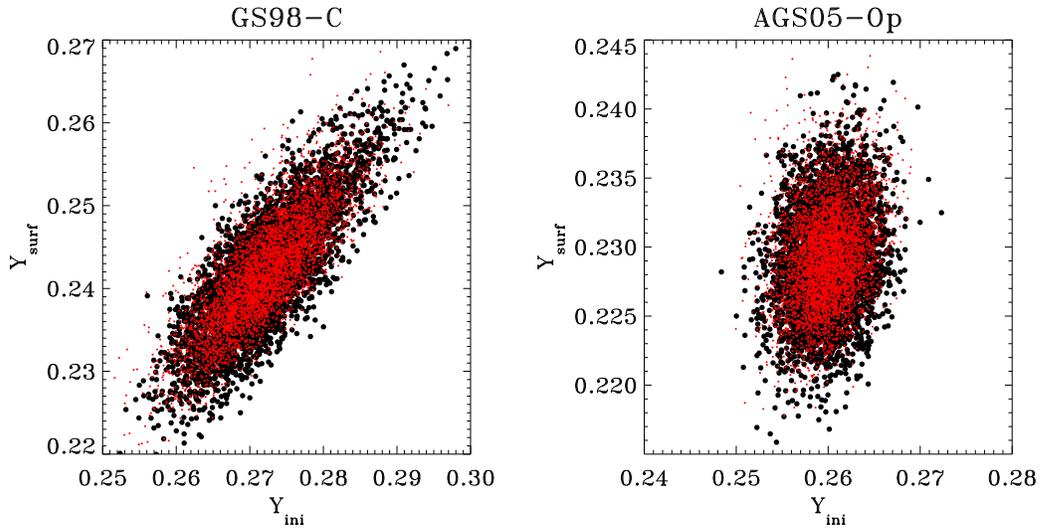}
\caption{Comparison between $\ysm$ vs. $\yim$ relation obtained with full
solar models (big black dots) with the obtained using power-law expressions
given in  Eqs.~\ref{eq:yini}  and  \ref{eq:ys}  and Table~\ref{tab:pwl} (small
red dots). Results for GS98 composition and conservative
uncertainties and AGS05 composition with optimistic uncertainties are
shown.\label{fig:ys-yini}} 
\end{figure}

We have verified  that Equations~\ref{eq:yini}~and~\ref{eq:ys} describe
the  behavior  of  $\yim$  and  $\ysm$  in solar  models very well  by 
comparing
distributions    generated   with these equations 
against results obtained with full  solar models.  In particular, we have used
the large 
set of solar  models computed for the Monte  Carlo simulation done by 
BSB06, where different assumptions for the reference solar
composition  and   abundance  uncertainties  were  adopted.   Results  of  the
comparison are presented in Figure~\ref{fig:ys-yini}.  The panel on the left shows 
results  for  the  GS98  composition  with  conservative  (large)
abundance uncertainties  (see BSB06 for  details); while the  panel on the right shows
results  for models using  AGS05 composition  and much smaller uncertainties, 
as originally given by  AGS05. In both cases, results from
the full  solar models obtained  from the BSB06  Monte Carlo simulations and
the resulting distributions  for  the $\yim$ and $\ysm$ obtained by means
of  Equations~\ref{eq:yini}~and~\ref{eq:ys} are shown.
We  see in  both  cases that  results from  full  solar models  are 
reproduced very closely by the power-law expressions. 

Having verified the validity of power-law expressions in reproducing the
correct distributions of $\yim$ and $\ysm$, in what follows, we concentrate on
using these expressions to constrain the solar initial abundance, $\yims$, in a
model independent manner. We do this by using the helioseismically determined
solar convection-zone helium abundance.

\section{CORRELATION BETWEEN $\ysm$ AND $\yim$}
\label{sec:correl}

In this  paper we  assume, unless specified  otherwise, that  only microscopic
diffusion of helium and heavy elements modify
the composition of the outer  layers of the Sun\footnote{The surface abundance
of lithium and beryllium are also modified by nuclear burning but 
that is not relevant  given the scope of the present work.}. Therefore,
in the absence of diffusion, $\ysm$ would exactly match $\yim$.
The most  evident effect of  diffusion, through gravitational settling,  is to
decrease $\ysm$ with respect to $\yim$. Thus $\ysm$ and
the  diffusion   rate  are   anticorrelated.   Diffusion  also   modifies  the
composition in the solar 
core. In particular,  gravitational settling of metals in the  core leads to a
higher radiative opacity  and hence, to a steeper  temperature profile. The
slightly  increased temperature  in turn increases the rates of nuclear
energy production. However, the  solar luminosity is fixed. Thus, the increase
in energy production  due to the larger temperatures has  to be compensated by
decreasing the  amount of available  hydrogen or, equivalently,  by increasing
the helium  abundance in the core. As a consequence,  $\yim$ is positively 
correlated with the
diffusion rate.  We call  these two combined  effects, the negative
correlation of $\ysm$  the the positive correlation 
of $\yim$ with the diffusion rate,  the `direct  influence' of
diffusion.

As can be seen from the coefficients listed in Table~\ref{tab:pwl}, the direct
influence of  diffusion is  the only effect  that
modify  $\ysm$ and $\yim$ in opposite ways.  For this  reason, in what  follows
we treat   the effect of diffusion  separately.   
Thus Equations.~\ref{eq:yini}~and~\ref{eq:ys} are modified to:

\begin{equation}
\left( \frac{\yim}{Y_{\rm ini,0}} \right)= \left(\frac{p_{\rm Diff}}{p_{\rm
    Diff,0}} \right)^{\alpha_{\rm Diff}} \left[ \prod_{i=1,N; i\neq {\rm Diff}} 
\left(\frac{p_i}{p_{i,0}} \right)^{\alpha_i} \right]
\label{eq:yi_2}
\end{equation}

\begin{equation}
\left(   \frac{\ysm}{Y_{\rm  surf,0}}   \right)=  \left(\frac{p_{\rm
    Diff}}{p_{\rm 
    Diff,0}} \right)^{\beta_{\rm Diff}} \left[\prod_{i=1,N; i\neq {\rm Diff}} 
\left(\frac{p_i}{p_{i,0}} \right)^{\beta_i} \right]
\label{eq:ys_2}
\end{equation}

\begin{figure}
\includegraphics[scale=0.9]{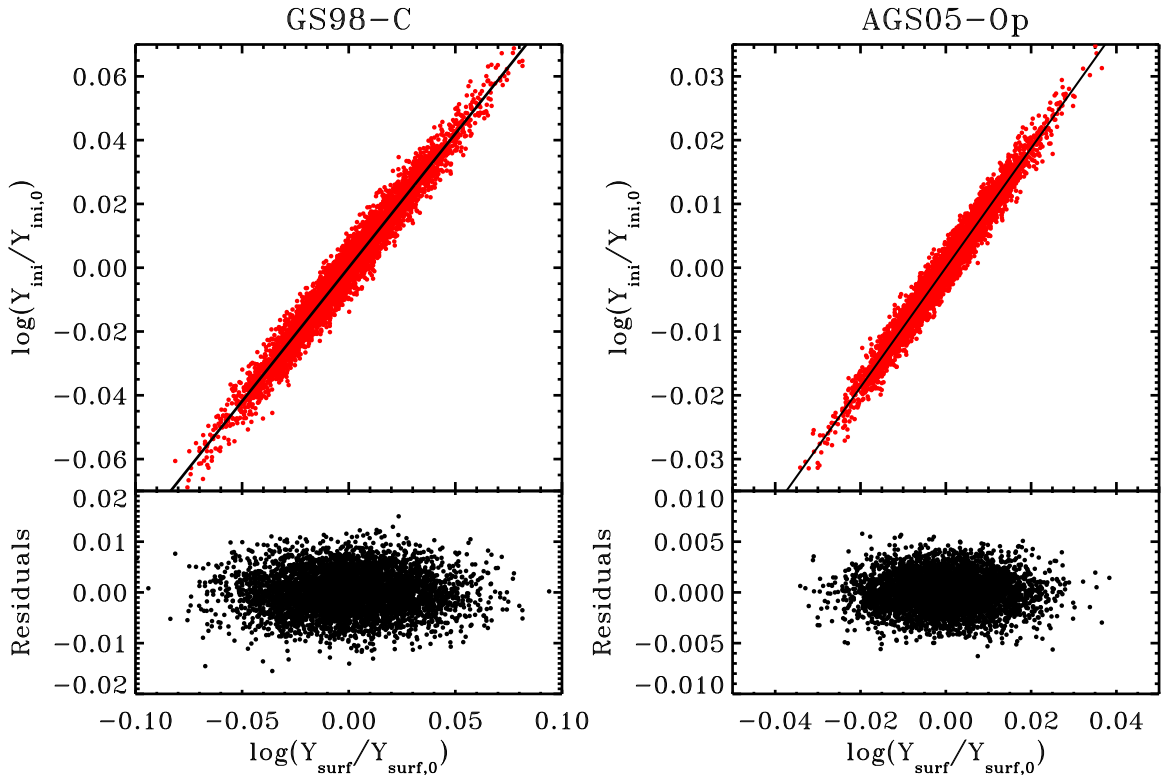}
\caption{$\yim$ vs. $\ysm$ distributions obtained using power-law expansions
(Equations~\ref{eq:yi_2} and~\ref{eq:ys_2})  but neglecting variations  in the
  diffusion rates (first factor in the equations). \label{fig:nodiff}}
\end{figure}

Using  the expressions in  brackets  in the  equations above,  we obtained
distributions for $\ysm$ and $\yim$ that exclude the direct effect of diffusion.
The    correlation   between these two quantities is    now    very    tight
as  can  be seen  in
Figure~\ref{fig:nodiff}. Naively,  we would expect  no dispersion to be
present, since  the  diffusion rates are kept  fixed. However, keeping the
diffusion rates fixed does not imply the total amount of helium (and
metals) diffused out of the convective envelope is constant.
The effective rate of  diffusion in the solar envelope is  directly related to the
extension  of   the  convective  envelope  which acts  as  a   reservoir  of
helium. Therefore, any changes to the  model inputs that have an effect on the
mass contained in  the convective envelope will be reflected in  a deviation of
the behavior of  $\ysm$ with respect to that  of $\yim$ (i.e., effectively
changing
the values of the  power-law exponents  $\alpha_i$ and  $\beta_i$ to a small
extent), leading to  the dispersion seen in Figure~\ref{fig:nodiff}.

Following a  procedure analogous to that used by \citet{haxton:2008} to study 
solar neutrino fluxes, we take advantage of the small dispersion in the
correlation between $\ysm$
and $\yim$  to express  the  uncertainties in  $\yim$ in  terms  of those  in
$\ysm$ 
while minimizing the  uncertainty in all  parameters except for diffusion. To do
this we combine Equations~\ref{eq:yi_2} and~\ref{eq:ys_2} to obtain: 
\begin{equation}
  \left(\frac{\yim}{Y_{\rm ini,0}}\right) = 
  \left(\frac{\ysm}{Y_{\rm surf,0}}\right)^{K_{\rm(I,S)}} 
\left(\frac{p_{\rm Diff}}{p_{\rm  Diff,0}} \right)^{\gamma_{\rm Diff}}
\left\{ \prod_{i=1,N; i\neq {\rm Diff}} 
\left(\frac{p_i}{p_{i,0}} \right)^{\gamma_i} \right\}
\end{equation}
where 
\begin{equation}
\gamma_i= \alpha_i - K_{\rm(I,S)} \beta_i.
\label{eq:gamma}
\end{equation}
The  constant $K_{\rm(I,S)}$  is  obtained by  doing a  linear fit  to the
simulated data for $\log({\yim})$ and $\log({\ysm})$ using the power-law 
expressions while ignoring the direct  dependence on diffusion. For the GS98-C
case we obtain $K_{\rm(I,S)}= 0.836$.  The residuals of the fit have a 
dispersion of 0.4\% (left panels of Figure~\ref{fig:nodiff}). For the exponent
in the diffusion factor we obtain $\gamma_{\rm Diff}=0.088$. 
We can now write
\begin{equation}
  \left(\frac{\yim}{Y_{\rm ini,0}}\right) = 
  \left(\frac{\ysm}{Y_{\rm  surf,0}}\right)^{0.836} \left(\frac{p_{\rm
      Diff}}{p_{\rm Diff,0}} \right)^{0.088} \left[1 \pm 0.4\%{\rm (ND)}
\right], 
\label{eq:first_res}
\end{equation}
where the uncertainty of 0.4\% represents the total uncertainty due to all 
non-diffusion  parameters  (note that  only  abundance-uncertainties play  an
appreciable role).  If we used optimistic abundance-uncertainties instead, the
total non-diffusion uncertainty would be only 0.2\%.

To proceed further, we need to consider the uncertainty of the 
diffusion rate (hereafter $\sigma_{\rm  Diff}$). 
The  best   agreement  between   the  structure  of   solar  models   and  the
helioseismically  determined  solar structure  is  obtained  when helium  and
heavy-element diffusion is included in the models \citep{bp:1995,modelS:1996}.
However, 
the  actual  diffusion  rates  of  helium  and  metals  and,  in
particular,     the     associated      uncertainties     are     not     well
constrained. \citet{tbl94} have estimated an uncertainty  of about 15\% in the 
diffusion rates. A similar result was obtained by \citet{fiore:1999} by using
conditions at the base of the convective zone to constrain the diffusion rates.
A detailed study by \citet{turcotte:1998} shows that diffusion velocities 
depend on the different assumptions made in the calculation for different
elements and, at different regions of the Sun, the variation ranges from a
few percent
to about 40\%. Here, we adopt an
intermediate fiducial value $\sigma_{\rm Diff}=20\%$ as the uncertainty in the
diffusion rates. It should be noted that the uncertainty in $\yim$
because of uncertainties in the diffusion rate are only one-twelfth that of 
uncertainties in the diffusion rate. Then, for $\sigma_{\rm Diff}=20\%$, the
contribution of diffusion to the total uncertainty in $\yim$  is only 1.7\%.
Thus,
\begin{equation}
  \left(\frac{\yim}{Y_{\rm ini,0}}\right) = 
 \left(\frac{\ysm}{Y_{\rm  surf,0}}\right)^{0.836}
\left[1 \pm 1.7\% {\rm (Diff)} \pm 0.4\%{\rm (ND)} \right].
\label{eq:final1}
\end{equation}

We now scale the relation to the Sun by adopting the 
present-day  surface helium  abundance determined
using helioseismology.  We adopt  the  value  determined  by
\citet{ys_basu04}, i.e., $\ysms= 0.2485\pm0.0034$, where the uncertainty
accounts for
systematics, including those caused by  uncertainties in the equation of state
(see \citealt{review_basu08} for a recent discussion about the determination of
$\ysms$). Then, the solar initial helium abundance $\yims$ can be expressed as 
\begin{equation}
\yims = Y_{\rm ini,0} 
  \left(\frac{0.2485}{Y_{\rm   surf,0}}\right)^{0.836}   \left[1  \pm
    1.1\%({\rm Helio}) \pm 1.7\%({\rm Diff}) \pm 0.4\%{\rm (ND)} \right]. 
\label{eq:final}
\end{equation}
In the above expression, standard solar models (SSMs) play a two-folded role. 
First, a  reference model is used to  get the  scaling factors  $Y_{\rm
ini,0}$ and $Y_{\rm surf,0}$.
Second, models are used to determine the exponent that relate $\ysms$  and
\ysssmp. However, as we show in the following section,
Equation~\ref{eq:final} has predictive power independent of the (standard)
solar model used as a reference.

\section{RESULTS}
\label{sec:results}

As a first test of Equation~\ref{eq:final}, we use values \yissm and \ysssm 
obtained from SSM calculations of \citet{ssm09}. The results are summarized in 
Table~\ref{tab:results} where we have identified the  solar models by the
solar 
composition adopted for each of them.  The second and third columns give the
 SSM predictions  for the initial and present-day surface
helium mass fraction.  The fourth column gives the $\yims$ values estimated
using Equation~\ref{eq:final}; the uncertainty 
in all cases is $\sigma_{\yims}=\pm 0.006$.  Comparing
results for the different standard 
solar models we find that the dispersion in $\yims$ is about ten times smaller
than $\sigma_{\yims}$ and hence we conclude that
the estimated values  of $\yims$ are consistent with each  other, and they are
independent of the standard solar model used as reference.

If  standard  solar  models  based   on  AGS05  composition are used to derive
the relation between $\yim$ and $\ysm$, we get $K_{\rm
  (I,S)}=0.903$  instead  of  0.836. We  have  used  GS98  and AGS05  for  this
comparison in  an attempt to bracket  what the real  solar abundance probably
is.  These two  compilations represent the two ends  of the currently accepted
range of values for solar photospheric abundances
\footnote{In  fact,   newer  determinations  of  solar abundances,  e.g.  by
\citet{agss09,caffau10}, point towards abundances of individual elements  in
the solar photosphere that are systematically higher than those in
\citet{ags05} (although consistent within  the quoted uncertainties), but lower
than those in \citet{gs98}.}. 
Therefore,  we   also   expect  that   our  estimates   of
$K_{\rm(I,S)}$ for the GS98-C and AGS05-Op cases represent the two extremes of
the possible range of values for this quantity. Replacing  
$K_{\rm (I,S)}=0.903$ in Eq.~\ref{eq:final} and  again using the standard
solar models from \citet{ssm09}, we  get  the  $\yims$  values listed 
in the last  column  of Table~\ref{tab:results}.  A comparison  of the
$\yims$ results
obtained with  the two values of $K_{\rm (I,S)}$ for each SSM
shows that the uncertainty in the determination of $K_{\rm (I,S)}$ has a
very minor effect in the  predicted $\yims$ values.  The largest difference is
found for the AGSS09ph model,  for which $\Delta \yims= 0.0017$, a
factor of $\sim 4$ smaller than $\sigma_{\yims}$ estimated above. 

We have  shown that the predicted $\yims$  values have a very  small dependence
on the solar model  used to determine it. The  choice of standard
solar models  for deriving  the  exponent $K_{\rm(I,S)}$  also  introduces a 
small scatter  in the  resulting  $\yims$ values.   Therefore,  we average 
all values $\yims$
 presented  in   Table~\ref{tab:results}  to   determine   the  final
result and also use the scatter to  define the systematic uncertainty of the method.
The final result that we obtain is
\begin{equation}
\yims= 0.278 \pm 0.006({\rm helio,param}) \pm 0.002({\rm syst})
\label{eq:res1}
\end{equation}
where  the   first  part  of  the  uncertainty   includes  contributions  from
helioseismology and solar input parameters, and the second one is the
systematic uncertainty of the method as defined above.

We have tested the robustness of this result by applying
Equation~(\ref{eq:final}) to a variety of solar models, standard and non-standard,
i.e. including additional physics not considered in SSMs.
As a first consistency check, we have applied Equation~\ref{eq:final} to the 
two large sets of SSMs computed by BSB06 used to assess the validity of the
power-law
expansions obtained in \S~\ref{sec:powerlaws}. Averaging all results for the
AGS05-Op Monte Carlo set presented in BSB06, we
find $\yims= 0.2780 \pm 0.0043$ and for the GS98-Cons set we get  $\yims= 0.2783 \pm
0.0041$,  i.e. results fully consistent with the basic derivation of $\yims$
and its uncertainty based on the four SSMs presented in
Table~\ref{tab:results}. These uncertainties do not include errors in the
helioseismic results.

To further test the robustness of our derivation of $\yims$ we have
resorted to results for standard and non-standard
solar models from the literature. Non-standard models
have been considered only when the modification to the input physics 
results in a model that is in better agreement with helioseismic results
than its SSM counterpart.
As a last test, we have computed additional
non-standard models to test the
influence of turbulent diffusion on the robustness of our predictions.

Here is a summary of our observations:
\begin{itemize}

\item[-] \citet{turcotte:1998} have tested the effect of different diffusion
schemes (including radiative  levitation in some cases) and the  use of
monochromatic opacities. By applying Equation~(\ref{eq:final}) to their models
that include 
diffusion, we obtain an average $\yims= 0.2780 \pm 0.0017$, where the
uncertainty refers to the scatter between models. From all their models we
infer $\yims$ consistent within $1\sigma$ of our result in
Equation~\ref{eq:res1}.

\item[-] Some standard and non-standard models have been presented in
\citet{bpb:01}. From these models (except their
models with no diffusion and with a mixed core; both clearly ruled out by
helioseismology) we determine an average value $\yims= 0.2774 \pm
0.0018$ (scatter between models). The model that most deviates from this
results includes rotation\footnote{According to \citet{pinsonneault:1999}, this
model sets a reasonable upper limit to rotation-induced mixing, compatible with
the observed lithium depletion in the Sun.}
which, by introducing additional mixing, tends to inhibit the efficiency of
diffusion. For this particular model we get
$\yims= 0.2725$, still within $1\sigma$ of the predicted value in
Equation~\ref{eq:res1}.

\item[-] \citet{booth:03} have computed a large set of different solar
models with different variations in the input physics and parameters. 
For all these models we derive $Y^{\rm ini}_\odot$ consistent within $1\sigma$
of the central result shown in Eq.~\ref{eq:res1}. 
The only exception is their model where the diffusion velocity of helium is
increased by 20\%, for which we derive $\ysms=0.2873$, a $1.5\sigma$ difference
with our derived central value for $\ysms$; but this model shows a
convective envelope too shallow compared to the helioseismic value, and hence,
we do not consider the discrepancy to be important. Averaging all results, we
get $Y^{\rm ini}_\odot=0.2783 \pm 0.0013$.

\item[-] We have computed solar models that include turbulent diffusion as given
by Equation~(2) of \citet{proffitt:1991}, where the diffusion coefficient
is $D_T=D_0 \left[\rho_{\rm CZ} / \rho\right]^3$, and $\rho$ is
the density and $\rho_{\rm CZ}$ the density at the base of the convective
envelope. As pointed out by \citet{jcd:2007}, even a small value for
$D_0$  (200 $\mathrm{cm^2  s^{-1}}$) eliminates  the bump  in  the sound-speed
difference between models and the Sun below the convective
envelope. We computed models for $D_0= 300, 1000,\ \mathrm{and},\ 3000\
\mathrm{cm^2 s^{-1}}$ with both GS98 and AGS05 compositions. Applying
Equation~(\ref{eq:final}) to these models, we  find $\yims= 0.2725,\ 0.2690,\
0.2665$ for increasing $D_0$ and irrespective of the composition adopted. 
It should be noted that all models with turbulent diffusion predict sound-speed
 profiles with average root-mean-square deviations from helioseimic results
that are comparable  to the rms deviations of SSMs. In  fact, for models based
on the GS98 composition we get  an average rms sound speed deviation of 0.0009
for the SSM while for models with $D_0= 300, 1000,\ \mathrm{and},\ 3000\
\mathrm{cm^2 s^{-1}}$  we get 0.0009, 0.0009, and  0.0010 respectively. For
models based on the AGS05 composition, results are 0.0047, 0.0053, 0.0054, and
0.0055 for the SSM and models with increasing $D_0$ as given above. 
This is a result of the fact that, although for models with turbulent diffusion the bump
below the CZ disappears, the lower central helium abundance in the
models lead to a slight worsening of the agreement with the
solar sound speed towards the center. This is also reflected in
the  frequency separation  ratios, particularly  $r_{02}$ \citep{bisoni:2007},
which in all three cases show worse agreement with the observed solar $r_{02}$
than the corresponding SSM constructed with the same composition. Larger
effects in this direction are found for increasing $D_0$ values. 
From  these results, we  conclude that  non-SSMs accounting  for the  effect of
extra mixing lead to the prediction:
\begin{equation}
\yims= 0.273 \pm 0.006({\rm helio,param}) \pm 0.002({\rm syst}),
\label{eq:res2}
\end{equation}
consistent with the result from SSMs given in \ref{eq:res2}.

\end{itemize}

\section{SUMMARY AND CONCLUSIONS}
\label{sec:summary}

We have used standard solar models to study the dependence of the initial
($\yim$) and surface ($\ysm$) helium abundances on different input physics and
element abundances entering solar model calculations. As expected, there is a
very tight correlation between these two quantities, particularly if direct
effects of diffusion are not taken into account. We have used this tight
correlation to express the initial solar helium abundance ($\yims$; not to be
confused with the initial value predicted by models) as a function of the
solar surface helium abundance,  $\ysms$, determined from helioseismology, while
minimizing the uncertainties from solar model input parameters. This result is 
expressed in a compact form in Equation~\ref{eq:final}. A slightly more general
expression where the scaling of the helium and metals diffusion rate is still
explicit is given in Equation~\ref{eq:first_res}.

We have applied Equation~(\ref{eq:final}) to different, recently
published standard solar models listed in Table~\ref{tab:results}, and derive
$\yims=0.278 \pm 0.006$ (Equation~\ref{eq:res1}). The uncertainty is dominated
by uncertainties in the diffusion rate of elements and those in the helioseismic
determination of  the solar  surface helium abundance.  The dispersion  in the
results for the different solar models is one order of magnitude smaller. The results
are also robust in terms of the standard solar models used to derive
Equation~\ref{eq:final}. 

To test further the robustness of our results, we have compiled results from
literature for standard and non-standard solar models (though only those
non-standard solar models that improve agreement with helioseismic
inferences of solar structure). In all cases, even for the non-standard ones,
applying Equation~\ref{eq:final} to results from solar models yield
$\yims$ results consistent to within $1\sigma$ of our central value  of
$\yims=0.278 \pm 0.006$.

Finally, we have also computed solar models including turbulent diffusion as a
phenomenological approach to eliminate the bump in the sound-speed difference between
 models and the Sun that is seen   below
the convective envelope. For these models, a small value of the free parameter $D_0$,
$D_0=300\, \mathrm{cm^2s^{-1}}$, leads to $\yims=0.2725$, consistent within
$1\sigma$ with our prediction given in Equation~\ref{eq:res1}. Larger values of
$D_0$ predict systematically lower values for $\yims$ but lead to slightly
degraded agreement with helioseismology, as evinced by frequency separation
ratios (probably a result of the  lower central helium abundance predicted by
these models). Thus, we find no evidence, at least from helioseismology data,
that support the need of larger $D_0$ values. It has to be kept in mind that 
this formulation of turbulent diffusion represents only a phenomenological
approach to eliminating the bump in the sound-speed difference profile. More
realistic 
models that account for rotationally-induced mixing, like that one derived by
\citet{pinsonneault:1999}, lead again to $\yims=0.2725$.

We conclude that the initial solar helium
abundance, as inferred from the present-day solar surface helium abundance
determined by helioseismology, is $\yims=0.278
\pm 0.006$. Although solar models are used to reach this conclusion, we find
this result is almost independent of which solar models are used in its
derivation. If $\yims$ is determined from non-standard solar models that account
for the effects of extra mixing, we obtain $\yims = 0.273\pm 0.006$. In all cases,
i.e., for both standard and non-standard models, we infer $\yims$ values that are higher than
the initial helium abundance obtained in solar models that use the  solar
abundances by \citet{ags05} or the most recent determination by \citet{agss09}.
This result points towards  a deficit in solar models using these abundances;
whether the problem lies in the abundance determinations or in the constitutive
physics of the models is beyond the scope of this work.

Depending on the solar metalicity adopted, we derive a helium-to-metal
enrichment of $\Delta Y/\Delta Z \sim 1.7-2.2$, in line with standard
predictions of Galactic chemical evolution and other determinations of $\Delta
Y/\Delta Z$ available in the literature.

\acknowledgements This work is partly supported by NSF grant ATM-0348837 to SB.

%\bibliographystyle{apj}
%\bibliography{helium}

\newpage 

\newpage 

\newpage 

\newpage 

\begin{deluxetable}{lr@{.}lr@{.}l|r@{.}lr@{.}l}
\tablewidth{0pt}
\tablecaption{Power-law  exponents for solar model  input parameters.
 \label{tab:pwl}} 
\tablehead{\colhead{ }  & \multicolumn{4}{c}{GS98} &
\multicolumn{4}{c}{AGS05} \\ \cline{2-9} & \multicolumn{2}{c}{$\yim$} & 
\multicolumn{2}{c}{$\ysm$}  &  \multicolumn{2}{c}{$\yim$} & 
\multicolumn{2}{c}{$\ysm$}
\\ 
 Solar param. & \multicolumn{2}{c}{$\alpha_i$} & \multicolumn{2}{c}{$\beta_i$} &
 \multicolumn{2}{c}{$\alpha_i$} & \multicolumn{2}{c}{$\beta_i$} }
\startdata
s11 & 0&0606 & 0&1290 & 0&0603 & 0&1393 \\ 
  s33 & -0&0037 & -0&0047 & -0&0034 & -0&0044 \\ 
  s34 & 0&0079 & 0&0098 & 0&0075 & 0&0095 \\ 
  s17 & 0&0000 & 0&0000 & 0&0000 & 0&0000 \\ 
sbe7e & 0&0000 & 0&0000 & 0&0000 & 0&0000 \\
 shep & 0&0000 & 0&0000 & 0&0000 & 0&0000 \\
sn14p & 0&0010 & 0&0011 & 0&0006 & 0&0007 \\
  age & -0&1360 & -0&1954 & -0&1408 & -0&2002 \\
 difu & 0&0271 & -0&0733 & 0&0331 & -0&0733 \\
 lumi & 0&3770 & 0&3522 & 0&4070 & 0&3810 \\
\hline C & -0&0079 & -0&0073 & -0&0040 & -0&0020 \\
N & -0&0017 & -0&0009 & -0&0004 & 0&0010 \\
O & 0&0083 & 0&0183 & 0&0140 & 0&0285 \\
Ne & 0&0285 & 0&0355 & 0&0237 & 0&0313 \\
Mg & 0&0277 & 0&0317 &0&0314 & 0&0366 \\
Si & 0&0593 & 0&0582 & 0&0682 & 0&0696 \\
S & 0&0439 & 0&0414 & 0&0506 & 0&0480 \\
Ar & 0&0123 & 0&0112 & 0&0075 & 0&0070 \\ 
Fe & 0&0813 & 0&0842 & 0&0877 & 0&0943 \\
\enddata
\end{deluxetable}

\begin{deluxetable}{lcccc}
\tablewidth{0pt}
\tablecaption{Initial solar helium abundance predictions from different standard
solar models}
\tablehead{
\colhead{\begin{tabular}{c}SSM\\ \ \end{tabular}} &
\colhead{\begin{tabular}{c}\yissm\\ \ \end{tabular}} &
\colhead{\begin{tabular}{c}\ysssm\\ \ \end{tabular}} &
\colhead{\begin{tabular}{c}$\yims$ \\ $K_{\rm(I,S)}=0.836$  \end{tabular}} &
\colhead{\begin{tabular}{c}$\yims$ \\ $K_{\rm(I,S)}=0.903$ \end{tabular}}}
\startdata
GS98 & 0.2721 & 0.2423 & 0.2779 & 0.2782 \\
AGS05 & 0.2593 & 0.2292 & 0.2774 & 0.2785 \\
AGSS09 & 0.2617 & 0.2314 & 0.2777 & 0.2787 \\
AGSS09ph & 0.2653 & 0.2349 & 0.2771 & 0.2788 \\
\enddata
\label{tab:results}
\end{deluxetable}

\end{document}